# A Definition of General Weighted Fairness and its Support in Explicit Rate Switch Algorithms [1] [2]

Bobby Vandalore, Sonia Fahmy, Raj Jain, Rohit Goyal, Mukul Goyal
The Ohio State University,
Department of Computer and Information Science
Columbus, OH 43210-1277
Phone: 614-292-3989, Fax: 614-292-2911
Email: {*vandalor, fahmy, jain, goyal, mukul*}@cis.ohio-state.edu

**Abstract**

In this paper we give a general definition of weighted fairness and show how this can achieve various fairness definitions, such as those mentioned in the ATM Forum TM 4.0 Specifications [1]. We discuss how a pricing policy can be mapped to general weighted (GW) fairness. The GW fairness can be achieved by calculating the *ExcessFairshare* (weighted fairshare of the left over bandwidth) for each VC. We show how a switch algorithm can be modified to support the GW fairness by using the *ExcessFairshare*. We use ERICA+ as an example switch algorithm and show how it can be modified to achieve the general fairness. Simulations results are presented to demonstrate that the modified switch algorithm achieves GW fairness. An analytical proof for convergence of the modified ERICA+ algorithm is given in the appendix.

# 1 Introduction

To guarantee a minimum amount of service the user can specify a MCR (minimum cell rate) in ABR service. The ABR service gives guarantee that the ACR (allowed cell rate) is never less than MCR. When MCR is zero for all sources, the available bandwidth can be allocated equally among the competing sources. This allocation achieves max-min fairness. When MCRs are non-zero, other definitions of fairness allocate the excess bandwidth (which is available ABR capacity less the sum of MCRs) equally among sources, or proportional to MCRs, or proportional to a predetermined weight assigned for different sources.

In the real world, the users prefer to get a service which reflects the amount they are paying. The pricing policy requirements can be realized by mapping appropriately the weights associated with the sources.

The specification of the ABR feedback control algorithm (switch algorithm) is not yet standardized. The earliest algorithms used binary feedback techniques [2]. Distributed algorithms [22] which emulated the centralized algorithm were proposed in [3, 4]. Improved, simpler distributed algorithms which achieved max-min fairness were proposed in [5, 6, 16, 18, 19, 20]. All the above algorithms assumed MCRs to be zero. Recently, [10, 11] discuss a generalized definition of max-min fairness and its distributed implementation. [12] discusses a weight-based max-min fairness policy and its implementation in ABR service. [13, 14] discuss the fairness in the presence of MCR guarantees.

In this paper we generalize the definition of the fairness, by allocating the excess bandwidth proportional to weights associated with each source. We show how a switch schemes can support non-zero MCRs and achieve the GW fairness. As an example, we show how the ERICA+ switch scheme can be modified to support GW fairness.

The modified scheme is tested using simulations on various configurations. The simulations test the performance of the modified algorithm, using different weights using a simple configurations, with transient sources, a link bottleneck configuration, and a source bottlenecked configuration. The simulations show that the scheme realizes various fairness definitions in ATM TM 4.0 specifications, which are special cases of the generalized fairness. We present an analytical proof of convergence for the modified algorithm in the appendix.

---

## 2 General Weighted Fairness: Definition

Define the following parameters:

$A_l$ = Total available bandwidth for all ABR connections on a given link $l$.

$A_b$ = Sum of bandwidth of underloaded connections which are bottlenecked elsewhere.

$A = A_l - A_b$, excess bandwidth, to be shared by connections bottlenecked on this link.

$N_a$ = Number of active connections

$N_b$ = Number of active connections bottlenecked elsewhere.

$n = N_a - N_b$, number of active connections bottlenecked on this link.

$\mu_i$ = MCR of connection $i$.

$\mu = \sum_{i=1}^{n} \mu_i$ Sum of MCRs of active connections within bottlenecked on this link.

$w_i$ = preassigned weight associated with the connection $i$.

$g_i$ = GW fair Allocation for connection $i$.

The general weighted fair allocation is defined as follows:

$$g_i = \mu_i + \frac{w_i(A - \mu)}{\sum_{j=1}^{n} w_j}$$

Note that this definition of fairness is different from the weighted allocation given as an example fairness criterion in ATM TM 4.0 specifications. In the above definition, only the excess bandwidth is allocated proportional to weights. This above definition ensures the allocation is at least MCR.

### 2.1 Mapping TM 4.0 Fairness to General Weighted Fairness

Here we show how the different fairness criteria mentioned in ATM TM 4.0 specification, can be realized based on the above fairness.

1. **Max-Min:** In this case MCRs are zero and the bandwidth is shared equally.

   $$g_i = A/n$$

   This is a special case of general weighted fairness with $\mu_i = 0$, and $w_i = c$, where c is a constant.

2. **MCR plus equal share:** The excess bandwith is shared equally.

   $$g_i = \mu_i + (A - \mu)/n$$

   by assigning equal weights we achieve the above fairness.

3. **Proportional to MCR:** The allocation is proportional to its MCR.

   $$g_i = \frac{A \times \mu_i}{\mu} = \frac{(\mu + A - \mu)\mu_i}{\mu} = \mu_i + \frac{(A - \mu)\mu_i}{\mu}$$

   By assigning $w_i = \mu_i$ we can achieve the above fairness.

# 3  Relationship to Pricing/Charging Policies

Consider a very small interval $T$ of time. The charge $C$ that a customer pays for using a network during this interval is a function of the number of bits W that the network transported successfully:

$$C = f(W, R)$$

Where, $R = W/T$ is the average rate.

It is reasonable to assume that $f()$ is a non-decreasing function of $W$. That is, those sending more bits do not pay less. The function $f()$ should also be a non-increasing function of time $T$ or equivalently a non-decreasing function of rate $R$.

For economy of scale, it is important that the cost per bit does not increase as the number of bits goes up. That is, $C/W$ is a non-decreasing function of $W$.

Mathematically, we have three requirements:

$$\partial C/\partial W \geq 0$$
$$\partial C/\partial R \geq 0$$
$$\partial (C/W)/\partial W \leq 0$$

One simple function that satisfies all these requirements is:

$$C = c + wW + rR$$

Here, $c$ is the fixed cost per connection; $w$ is the cost per bit; and $r$ is the cost per Mbps. In general, $c$, $w$, and $r$ can take any non-negative value.

In the presence of MCR, the above discussion can be generalized to:

$$C = f(W, R, M)$$

Where, $M$ is the MCR. All arguments given above for $R$ apply to $M$ also except that the customers requesting larger $M$ possibly pay more. One possible function is:

$$C = c + wW + rR + mM$$

where, $m$ is dollars per Mbps of MCR. In effect, the customer pays $r + m$ dollars per Mbps up to $M$ and then pays only $r$ dollars per Mbps for all the extra bandwidth he/she gets over and above $M$.

Consider two users with MCRs $M_1$ and $M_2$. Suppose their allocated rates are $R_1$ and $R_2$ and, thus, they transmit $W_1$ and $W_2$ bits, respectively. Their costs are:

$$C_1 = c + wW_1 + rR_1 + mM_1$$
$$C_2 = c + wW_2 + rR_2 + mM_2$$

Cost per bit $(C/W)$ should be a decreasing function of bits $W$. Thus, if $W_1 \geq W_2$:

$$C_1/W_1 \leq C_2/W_2$$

$$c/W_1 + w + rR_1/W_1 + mM_1/W_1 \leq c/W_2 + w + rR_2/W_2 + mM_2/W_2$$

Since $R_i = W_i/T$, we have:

$$c/(R_1 T) + w + r/T + mM_1/(R_1 T) \leq c/(R_2 T) + w + r/T + mM_2/(R_2 T)$$

$$c/R_1 + mM_1/R_1 \leq c/R_2 + mM_2/R_2$$

$$(c + mM_1)/(c + mM_2) \leq R_1/R_2$$

$$(a + M_1)/(a + M_2) \leq R_1/R_2$$

Where $a$ $(=c/m)$ is the ratio of the fixed cost and cost per unit of MCR.

Note that the allocated rates should either be proportional to $a$+MCR or be a non-decreasing function of MCR. This is the weight policy we have chosen to use in our simulations.

## 4 General Weight Fair Allocation: Problem Formulation

In this section we give the formal specification of the general weighted fair allocation problem, and give a motivation for the need of a distributed algorithm.

The following additional notation are necessary:

$\mathcal{L}$ = Set of links, $\mathcal{L}_s$ set of links which session $s$ goes through.

$\mathcal{S}$ = Set of sessions, $\mathcal{S}_l$ set of sessions which go through link $l$. $N = |S|$.

$\mathcal{A} = (\mathcal{A}_l, l \in \mathcal{L})$ set of of available capacity.

$\mathcal{M} = (\mu_s, s \in S)$, where $\mu_s$ is the minimum cell rate (MCR) for session $s$.

$\mathcal{W} = (w_1, w_2, \ldots, w_N)$ denotes the weight vector.

$\mathcal{R} = (r_1, r_2, \ldots, r_N)$ the current allocation vector (or rate vector).

$\mathcal{G} = (g_1, g_2, \ldots, g_N)$ the general fair allocation. $\mathcal{G}_{\mathcal{S}_l}$ denotes the set of allocations of sessions going over link $l$

**Definition 1** *General Weighted Fair Allocation Problem*

*The GW fair problem is to find the rate vector equal to the GW fair allocation, i.e., $\mathcal{R} = \mathcal{G}$. Where $g_i \in \mathcal{G}_{\mathcal{S}_l}$ is calculated for each link $l$ as defined in the section 2.*

Note the 5-tuple $(\mathcal{S}, \mathcal{L}, \mathcal{C}, \mathcal{W}, \mathcal{R})$ represents an instant of the bandwidth sharing problem. When all weights are equal the allocation is equivalent to the general max-min fair allocation as defined in [10, 11]. A simple centralized algorithm for solving the above problem would be to first, find the correct allocation vector for the bottleneck links. Then, solve the same problem of smaller size after deleting bottleneck links. A similar kind of centralized, recursive algorithm is discussed in [10]. Centralized algorithm implies that all information is known at each switch, which is not feasible, hence a distributed algorithm is necessary.

# 5 Achieving General Fairness in Switch Algorithms

A typical ABR switch scheme calculates the excess bandwidth capacity available for best effort ABR after reserving bandwidth for providing MCR guarantee and higher priority classes such as VBR and CBR. The switch fairly divides the excess bandwidth among the connections bottlenecked at that link. Therefore, the ACR can be represented by the following equation.

$$ACR(i) = \mu_i + ExcessFairshare(i)$$

$ExcessFairshare$ is the amount of bandwidth allocated over the MCR in a fair manner.

In the case of GW fairness, the $ExcessFairshare$ term is given by:

$$ExcessFairshare(i) = \frac{w_i(A - \mu)}{\sum_{j=1}^{n} w_j}$$

If the network is near steady state (input rate = available capacity), then the above allocation enables the sources to attain the GW fairness. The ATM TM 4.0 specification mentions that the value of $(ACR - MCR)$ can be used in the switch algorithms, we use this term to achieve GW fairness. We have to ensure the $(ACR - MCR)$ converges to the $ExcessFairshare$. We use the notion of *activity level* to achieve the above [9]. A connection's *activity level* ($AL(i)$) is defined as follows.

$$AL(i) = minimum\left(1, \frac{SourceRate(i) - \mu_i}{ExcessFairshare(i)}\right)$$

$SourceRate(i)$ is the rate at which the source is currently transmitting data. Note that, $SourceRate(i)$ is the $ACR(i)$ given as the feedback rate earlier by the switch. The activity level indicates how much of the $ExcessFairshare$ is actually being used by the connection. The activity level attains the value of one when the $ExcessFairshare$ is used by the connection. It is interesting to note that using activity level for calculating is similar to the Charny's [15] *consistent marking* technique, where switch marks connections which have lower rate than their *advertised rate*. The new advertised rate is calculated using the equation:

$$\text{Advertised Rate} = \frac{\mathcal{A}_l - \sum \text{Rates of marked connections}}{\mid S_l \mid - \sum \text{Marked connections}}$$

The activity level inherently captures the notion of marking, i.e., when a source is bottlenecked elsewhere, then activity level times the fairshare (based on available left over capacity) is the actual fairshare of the bottleneck source. The computation of activity level can be done locally and is an $O(1)$ operation, compared to $O(n)$ computations required in consistent marking [15].

We expect that the links use their $ExcessFairshare$, but this might not be case. By multiplying the weights by the activity level, and using these as the weights in calculating the $ExcessFairshare$ we can make sure that the rates converge to the GW fairness allocation. Therefore, the $ExcessFairshare$ share term is defined as:

$$ExcessFairshare(i) = \frac{w_i AL(i)(A - \mu)}{\sum_{j=1}^{n} w_j AL(j)}$$

An switch algorithm can use the above $ExcessFairshare$ term to achieve general fairness. In the next section we show how the ERICA+ switching algorithm is modified to achieve GW fairness.

# 6　Example Modifications to A Switch Algorithm

The ERICA+ algorithm operates at each output port of a switch. The switch periodically monitors the load on each link and determines a load factor ($z$), the available ABR capacity, and number of currently active sources or VCs (N). The measurement period is the "Averaging Interval". These measurements are used to calculate the feedback rate which is indicated in the BRM (backward RM) cells. The measurements are done in the forward direction and the feedback is given int the backward direction. The complete description of the ERICA+ algorithm can be obtained from [5].

The ERICA+ algorithm uses the term $FairShare$ which is the bottleneck link capacity divided by the active number of VCs. It also uses $MaxAllocPrevious$ term, which is the maximum allocation in the previous "Averaging Interval". This term is used to achieve Max-min fairness. We modify the algorithm by replacing the $FairShare$ term by $ExcessFairshare(i)$ and adding the $\mu_i$. The keys steps in ERICA+ which are modified to achieve the GW fairness shown below:

**Algorithm A** At the end of Averaging Interval:

$$\text{Total ABR Capacity} \leftarrow \text{Link Capacity} - \text{VBR Capacity}$$
$$- \sum_{i=0}^{n} \min(SourceRate(i), \mu_i) \tag{1}$$

$$\text{Target ABR Capacity} \leftarrow Fraction \times \text{Total ABR Capacity} \tag{2}$$

$$\text{Input Rate} \leftarrow \text{ABR Input Rate} - \sum_{i=0}^{n} \min(SourceRate(i), \mu_i) \tag{3}$$

$$z \leftarrow \frac{\text{Input Rate}}{\text{Target ABR Capacity}} \tag{4}$$

$$ExcessFairshare(i) \leftarrow \frac{(\text{Target ABR Capacity}) w_i AL(i)}{\sum_{j=1}^{n} w_j AL(j)} \tag{5}$$

$$\tag{6}$$

The $Fraction$ term is dependent on the queue length [8]. Its value is one for small queue lengths and drops sharply as queue length increases. When the $Fraction$ is less than one, $(1 - Fraction) \times TotalABRCapacity$ is used to drain the queues. ERICA+ uses an hyperbolic function calculating value of the $Fraction$.

**When a BRM is received:**

$$VCShare \leftarrow \frac{SourceRate(i) - \mu_i}{z} \tag{7}$$

$$ER \leftarrow \mu_i + \text{Max}(ExcessFairshare(i), VCShare) \tag{8}$$

$$ER\_in\_RM\_Cell \leftarrow \text{Min}(ER\_in\_RM\_Cell, ER, \text{Target ABR Capacity}) \tag{9}$$

The $VCShare$ is used to achieve an unit overload. When the network reaches steady state the $VCShare$ term converges to $ExcessFairshare(i)$ term achieving generalized fairness criterion. The complexity of the computations done at the switching interval is $O(number of VCs)$. The update operation when the BRM cell arrives is an $O(1)$ operation. Proof of convergence of algorithm A, is given in the appendix.

# 7　Simulation Configurations

We use different configurations to test the performance of the modified algorithm. We assume that the sources are greedy, i.e., they have infinite amount of data to send, and always send data at ACR. In all configurations the data traffic is only one way, from source to destination. All the link bandwidths are 149.76 (155.52 less the SONET overhead), expect in the GFC-2 configuration.

## 7.1 Three Sources

This is a simple configuration in which three sources send data to three destinations over a two switchs and a bottleneck link. See figure 1. Only sources send data. This configuration is used to demonstrate that the modified switchs algorithm can achieve the general fairness for the various set of weight assignments.

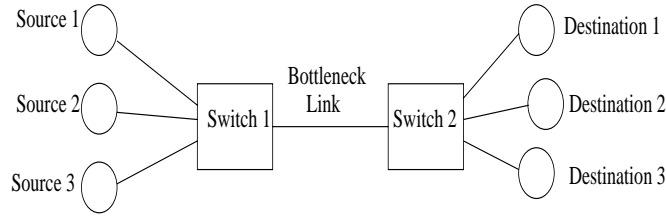

Figure 1: N Sources - N Destinations Configuration

## 7.2 Source Bottleneck

In this configuration, the source S1, is bottlenecked to rate (10 Mbps), which below its fairshare (50 Mbps). This configuration tests whether the fairness criterion can be achieved in the presence of source bottleneck.

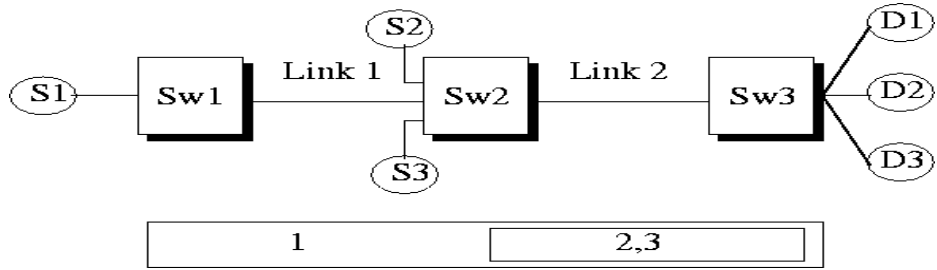

Figure 2: 3 Sources - Bottleneck Configuration

## 7.3 Generic Fairness Configuration - 2 (GFC-2)

This configuration is a combination of upstream and parking lot configuration (See Figure 3). In the configuration all the links are bottlenecked links. This configuration is explained in [7].

## 7.4 Simulation Parameters

The simulations were done on extensively modified version of NIST simulator [17]. The parameters values used in the different configurations is given in Table 1. The "Averaging Interval" is the period for which the switch monitors various parameters. Feedback is given based on these monitored values. The ERICA+ algorithm uses dynamic queue control to vary the available ABR capacity dependent on queue size. At steady state the queue length remains constant. The "Target Delay" parameter specifies the desired delay due to this constant queue length at steady state.

# 8 Simulation Results

In this section we give the simulation results for the different configurations.

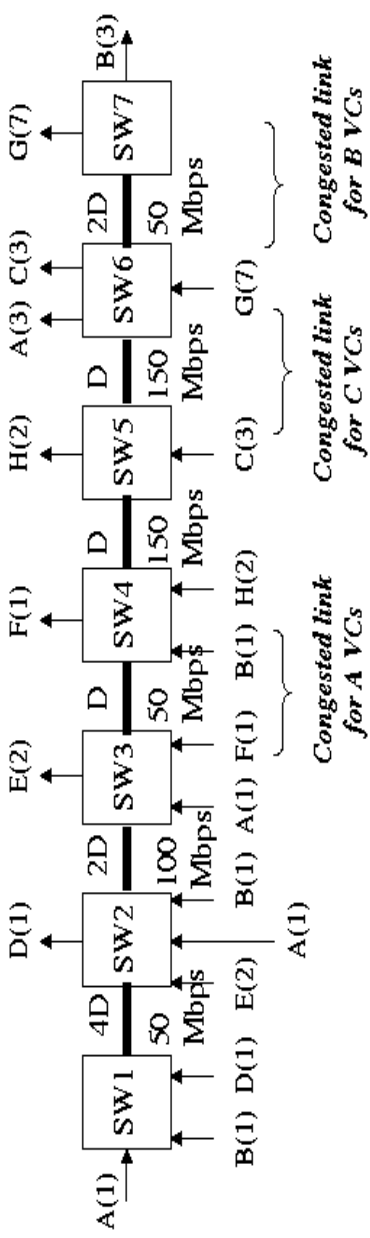

Figure 3: Generic Fairness Configuration - 2

Table 1: Simulation Parameter Values

| Configuration Name | Link Distance | Averaging interval | Target Delay |
|---|---|---|---|
| Three Sources | 1000 Km | 5 ms | 1.5 ms |
| Source Bottleneck | 1000 Km | 5 ms | 1.5 ms |
| GFC-2 | 1000 Km | 15 ms | 1.5 ms |

## 8.1 Three Sources

Simulations using a number of weight functions were done using the simple three sources configuration to demonstrate that general fairness is achieved in all these cases. The ICRs (initial cell rate) of the sources were set to the (50,40,55) in all the simulations.

The results of these cases are given in Table 2. The following can be observed from the Table 2

- Case 1: $a = \infty$, MCRs = 0. All weights are equal so the allocation (149.76/3) = 49.92 for each connection. This is allocation is the same as max-min fair allocation.

- Case 2: $a = \infty$, MCRs $\neq$ 0. The left over capacity 149.76 - (10 + 30 + 50) = 59.76 is divided equally among the three sources. So the allocation is (10 + 19.92, 30 + 19.92, 50 + 19.92) = (29.92,39.92,69.92).

- Case 3: $a = 5$, MCRs $\neq$ 1. Hence, the weight function is 5 + MCR. The left over capacity 59.76 Mbps, is divided proportional to (15,35,55). Hence the allocation is (10 + 15/105 × 59.76, 30 + 35/105 × 59.76, 50 + 55/105 × 59.76) = (16.64, 49.92, 83.2).

The Figure 2 shows the ACRs of the three sources and the queue length to bottleneck link at switch-1 for the above three cases. From the figure one can observe that the sources achieve the GW fairness and the queue length is constant in steady state.

Table 2: Three sources configuration simulation results

| Case Number | Src Num | mcr | a | weight function | Expected fair share | Actual share |
|---|---|---|---|---|---|---|
| 1 | 1 | 0 | $\infty$ | 1 | 49.92 | 49.92 |
|   | 2 | 0 | $\infty$ | 1 | 49.92 | 49.92 |
|   | 3 | 0 | $\infty$ | 1 | 49.92 | 49.92 |
| 2 | 1 | 10 | $\infty$ | 1 | 29.92 | 29.92 |
|   | 2 | 30 | $\infty$ | 1 | 49.92 | 49.92 |
|   | 3 | 50 | $\infty$ | 1 | 69.92 | 69.92 |
| 3 | 1 | 10 | 5 | 15 | 18.53 | 16.64 |
|   | 2 | 30 | 5 | 35 | 49.92 | 49.92 |
|   | 3 | 50 | 5 | 55 | 81.30 | 81.30 |

Table 3: Three sources transient configuration simulation results

| Case Number | Src Num | mcr | a | weight function | Expected fairshare (non-trans.) | Actual (non-trans) share | Expected fairshare (trans.) | Actual (trans.) share |
|---|---|---|---|---|---|---|---|---|
| 1 | 1 | 0 | $\infty$ | 1 | 74.88 | 74.83 | 49.92 | 49.92 |
|   | 2 | 0 | $\infty$ | 1 | NC | NC | 49.92 | 49.92 |
|   | 3 | 0 | $\infty$ | 1 | 74.88 | 74.83 | 49.92 | 49.92 |
| 2 | 1 | 10 | $\infty$ | 1 | 54.88 | 54.88 | 29.92 | 29.83 |
|   | 2 | 30 | $\infty$ | 1 | NC | NC | 49.92 | 49.92 |
|   | 3 | 50 | $\infty$ | 1 | 94.88 | 95.81 | 69.92 | 70.93 |
| 3 | 1 | 10 | 5 | 15 | 29.92 | 29.23 | 18.53 | 18.53 |
|   | 2 | 30 | 5 | 35 | NC | NC | 49.92 | 49.92 |
|   | 3 | 50 | 5 | 55 | 119.84 | 120.71 | 81.30 | 81.94 |

NC - not converged

## 8.2 Three Sources: Transient

In these simulations the same simple three source configuration is used. The source 1 and source 3 transmit data throughout the simulation period. The source 2 is a transient source, which starts transmitting at 400 ms and stops at 800 ms. The total simulation time is 1200 ms. The same parameters values from the case's 1, 2 and 3 of the previous section were used in these simulations. The results of these simulations are given in Table 3. The (non-trans.) columns give the allocation when transient source 2 is not present, i.e., between 0ms to 400ms and between 800ms to 1200 ms. The (trans.) columns give allocation when the transient source 2 is present, i.e., between 400 ms to 800 ms.

The graphs for these three simulations are shown in figure 5. The graphs show the ACRs of the three sources and the bottleneck link utilization. It can be seen both from the Table 3 and the graphs that the switch algorithm does converge to the general fairness allocation even in the presence of transient sources. Note the link utilization is high throughout the simulation. The width of dip in the utilization graph (Figures 5(b),(d),(f)) when the transient sources goes away, indicates the responsiveness of the algorithm. This shows that the algorithm is tolerant of transient sources and responds quickly to changing demands.

Table 4: Three sources bottleneck configuration simulation results

| Case Number | Src Num | mcr | a | weight function | Expected fair share | Using CCR in RM cell | Using Measured CCR |
|---|---|---|---|---|---|---|---|
| 1 | 1 | 0 | $\infty$ | 1 | 49.92 | 49.85 | 49.92 |
|   | 2 | 0 | $\infty$ | 1 | 49.92 | 49.92 | 49.92 |
|   | 3 | 0 | $\infty$ | 1 | 49.92 | 49.92 | 49.92 |
| 2 | 1 | 10 | $\infty$ | 1 | 29.92 | NC | 29.62 |
|   | 2 | 30 | $\infty$ | 1 | 49.92 | NC | 49.60 |
|   | 3 | 50 | $\infty$ | 1 | 69.92 | NC | 71.03 |
| 3 | 1 | 10 | 5 | 15 | 18.53 | NC | 18.42 |
|   | 2 | 30 | 5 | 35 | 49.92 | NC | 49.92 |
|   | 3 | 50 | 5 | 35 | 81.30 | NC | 81.93 |

NC - not converged

## 8.3 Source Bottleneck

The case 1, 2 and 3 of section 8.1 were simulated using the three sources bottleneck configuration. In these simulations the source S1 is bottlenecked at 10 Mbps, i.e., it always transmits data at rate of at most 10 Mbps, irrespective of its ACR (and ICR). The initial ICRs were set to 50, 30, 110. The load on the bottleneck link is near unity. If the switch algorithm uses the CCR (current cell rate) value indicated in the RM cell as the source rate the switch cannot estimate the correct value of source rate of the bottleneck source. But if the switch uses measured source rate then it can correctly estimate the bottlenecked source's rate. Table 4 shows the results both when the switch uses the CCR field and when it measure's the source rate. The correct fairness is achieved when the measured source rates are used.

The graphs for the simulations are given in Figure 6. The switch algorithm uses queue control, to dynamically use part of available ABR capacity to drain the queues. When the queue is large the available ABR capacity is only a fraction of actual capacity. So, the algorithm takes sometime before converging to the correct fairness values. When the CCR value from the RM cells is used, the algorithm is not able to estimate the actual rate at which the source is sending data. So it does not converge in case 2 and case 3 (Figures 6(c) and 6(e)). In case 1 (Figure 6(a), it converged since the bottleneck source's rate (CCR) had the correct value of 50 which is the same allocation it would get in the fair allocation.

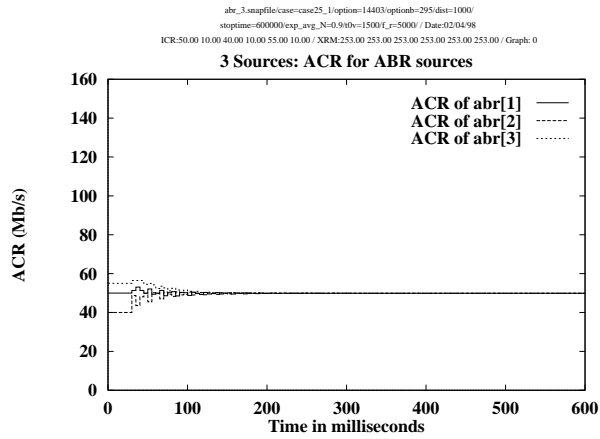
(a)

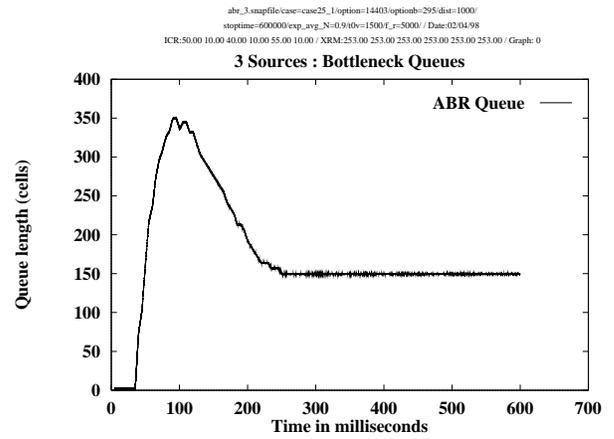
(b)

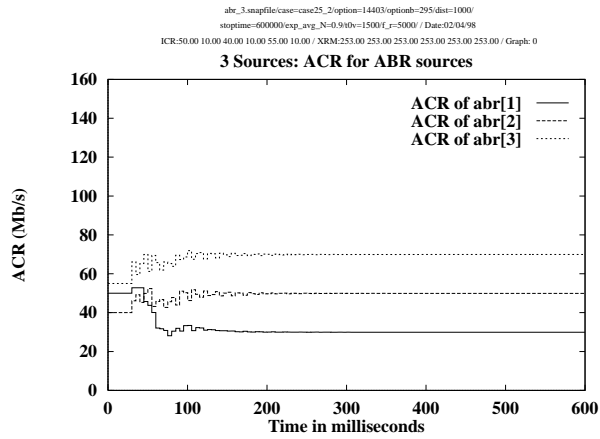
(c)

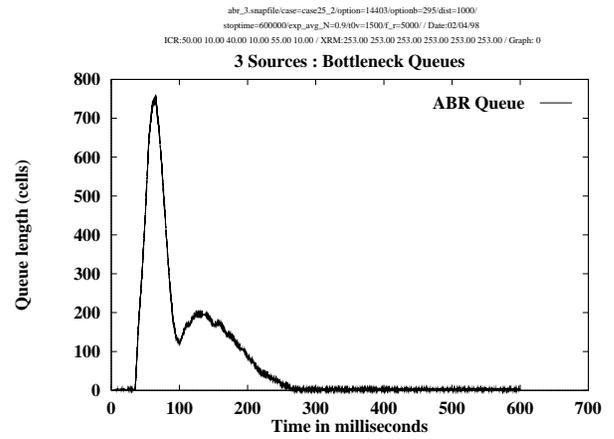
(d)

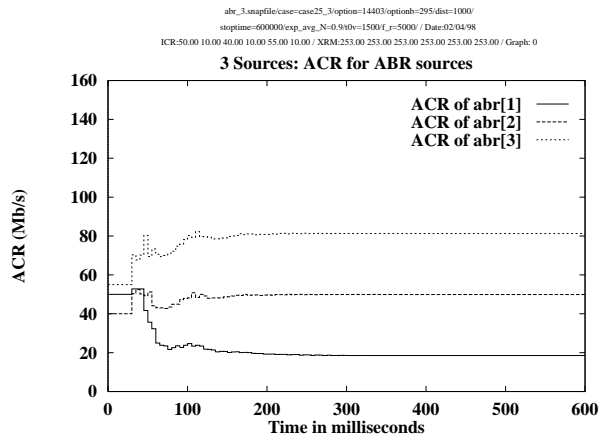
(e)

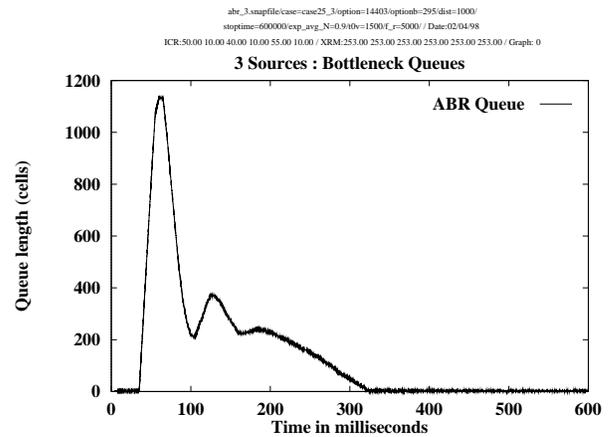
(f)

Figure 4: Three Sources: ACR graphs and Queue length of bottleneck link

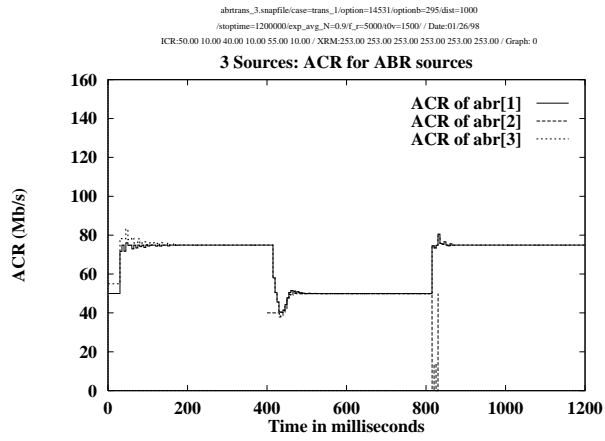

(a)

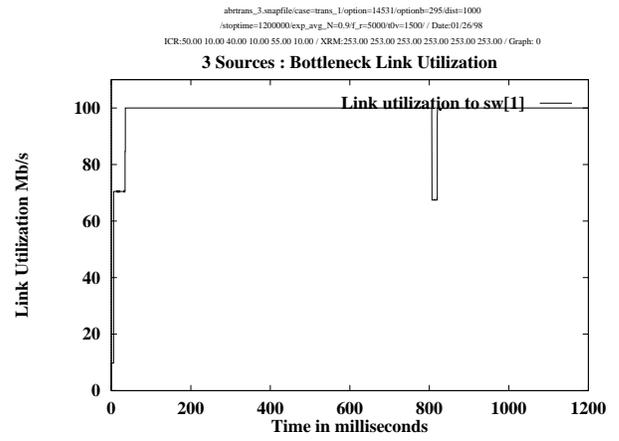

(b)

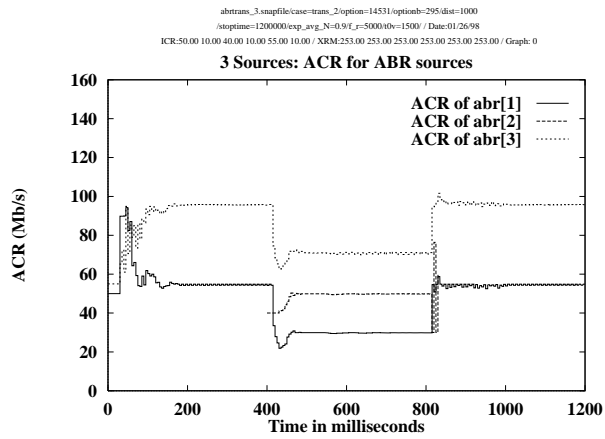

(c)

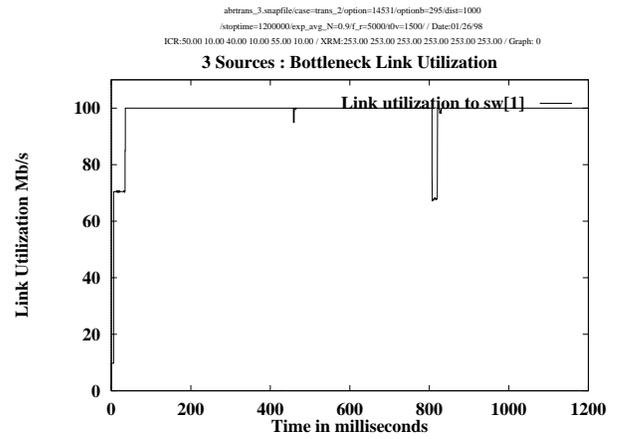

(d)

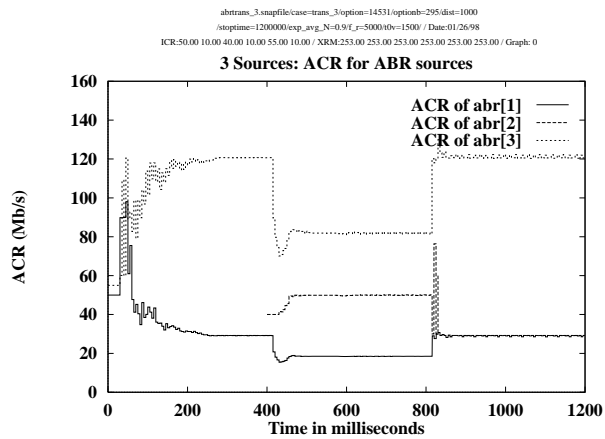

(e)

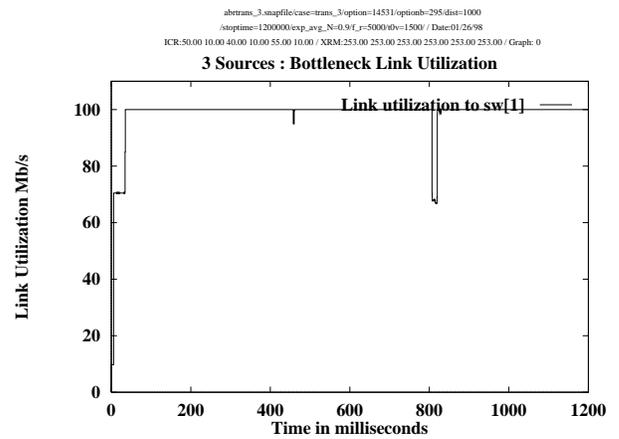

(f)

Figure 5: Three Sources (Transient) : ACR and Utilization graphs

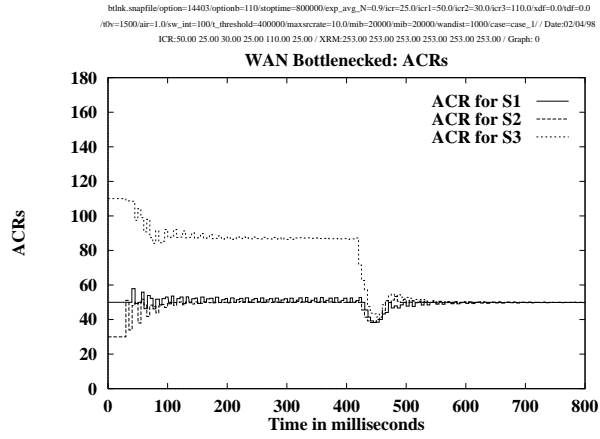
(a)
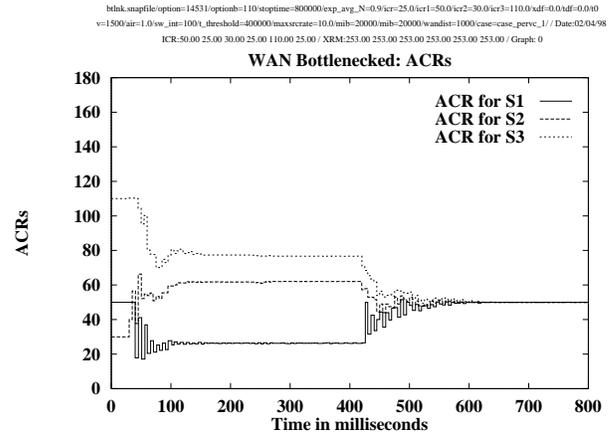
(b)

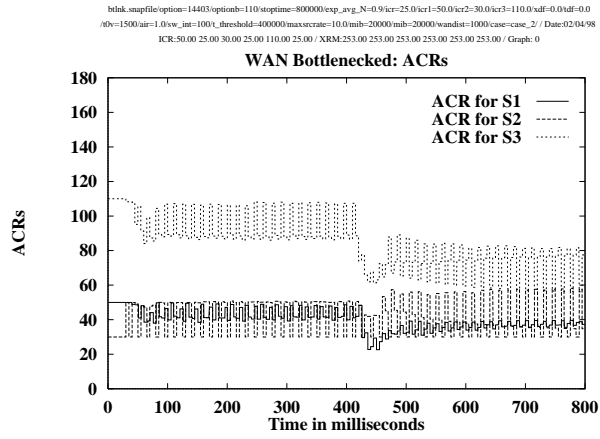
(c)
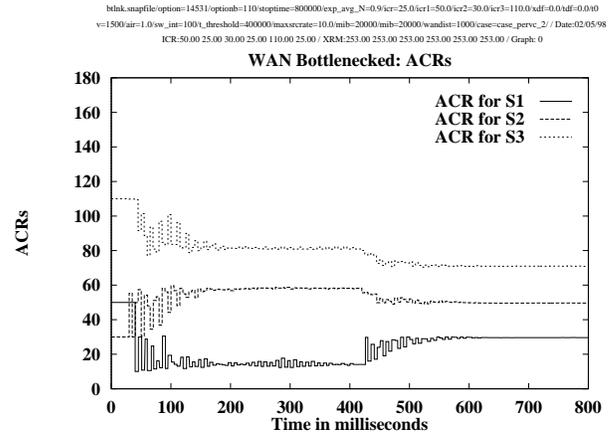
(d)

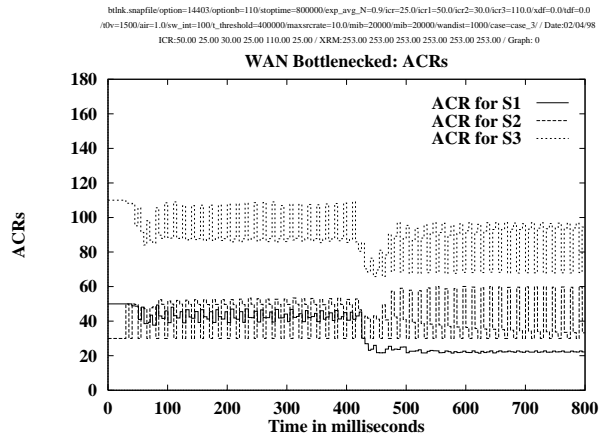
(e)
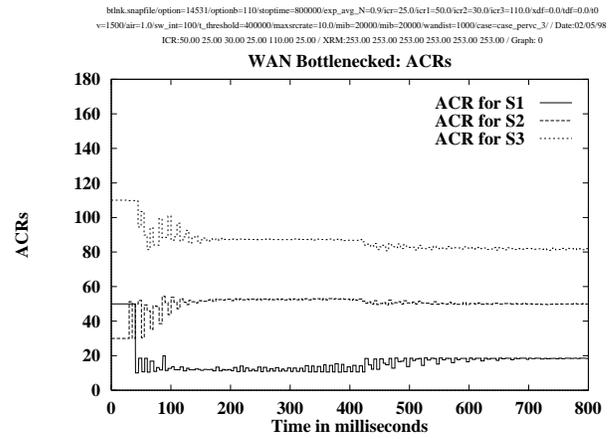
(f)

Figure 6: Three Sources Bottleneck: ACR graphs (with and without measuring Source Rate)

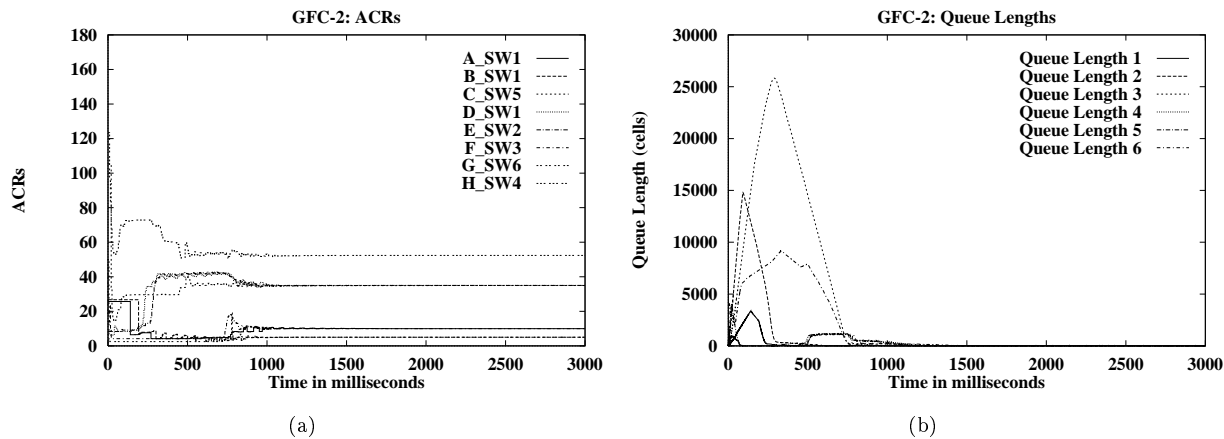

Figure 7: GFC-2 configuration: ACRs of A through H, VCs and Queue lengths at bottlenecks links

Table 5: GFC-2 configuration: simulation results

| Case Number | VC type | Expected allocation | Actual Allocation |
|---|---|---|---|
| 1 | A | 10 | 9.85 |
| | B | 5 | 4.97 |
| (a = ∞) | C | 35 | 35.56 |
| | D | 35 | 35.71 |
| (all MCRs | E | 35 | 35.34 |
| are zero) | F | 10 | 10.75 |
| (same as | G | 5 | 5.00 |
| max-min) | H | 52.5 | 51.95 |

### 8.4 Link Bottleneck: GFC-2

In this configuration each link is a bottleneck link. The Figure 7 (a) shows the ACR graphs for each type of VCs. Figure 7 (b) shows the queue length of all the bottleneck links (links between the switches). From the Figure and Table 5 it can be seen that the VCs converge to their expected fairshare. This shows that the algorithm works in the presence of link bottlenecks.

## 9 Conclusion

In this paper, we have given a general definition of fairness, which inherently provides MCR guarantee and divides the excess bandwidth proportional to predetermined weights. Different fairness criterion such as max-min fairness, MCR plus equal share, proportional MCR can be realized as special cases of this general fairness. We showed how to realize a typical pricing policy by appropriate weight function. The general fairness can be achieved by using the *ExcessFairshare* term in the switch algorithms. The weights are multiplied by the activity level when calculating the *ExcessFairshare* to reflect the actual usage of the source.

We have shown how ERICA+ switch algorithm can be modified achieve this general fairness. The proof of convergence of algorithm A is given in the appendix. The modified algorithm has been tested under different configuration using persistent sources. The simulations results show that the modified algorithm achieves the general fairness in all configurations. In addition, the results show that the algorithm converges in the presence of both source and link bottleneck and is quick to respond in the presence of transient sources. In source bottlenecked configuration the value of the CCR (source rate) from the RM cell maybe incorrect. Hence, it is necessary to used the measured source rate in the presence of source bottlenecks.

# Appendix: Proof of convergence of Algorithm A

We make the following assumptions:

- Synchronous update of source rates
- Queue control function is a constant function
- Infinite (greedy) sources, which always have data to send. Though there might be source or link bottleneck present.
- If a source bottleneck is present, it does not change it bottleneck rate during convergence.
- $\sum_{s \in S_l} \mu_i \leq A_l$
- Load factor $z > 0$ and $ER < A_l < LinkRate$

**Lemma 1** *The Algorithm A converges to the GW fair allocation, for a sessions bottlenecked by a link.*

**Proof:** The proof technique used here is similar to the one used in [5]. Let $l_b$ be the link which is bottlenecked. Without loss of generality assume that first $k$ sessions through the link $l_b$ are bottlenecked (either link bottlenecked or source bottlenecked) elsewhere. Let $n = \mid S_{l_b} \mid -k$. Let $r_{b1}, r_{b2}, \ldots, r_{bk}$ be the bottleneck rates and $r_1, r_2, \ldots, r_n$ be the rates of non-bottlenecked (underloaded) sources. Let $A_b = \sum_{i=1}^{k} r_{bi}$ be total capacity of bottlenecked links. These non-bottlenecked sources are bottlenecked by the current link $l_b$. According to the definition of the general fair allocation the rates $g_i$ is given by:

$$g_i = \mu_i + \frac{w_i(A_l - A_b)}{\sum_{j=1}^{n} w_j}$$

Assume that the bottlenecks elsewhere have been achieved, there for the rates $r_{b1}, r_{b2}, \ldots, r_{bk}$ are stable. For simplicity, assume that the MCRs of these sources are zero. Proof for the bottlenecks having non-zero MCRs is a simple extension.

We show that rates allocated at this switch converges to $r_{b1}, r_{b2}, \ldots, r_{bk}$ and $g_1, g_2, \ldots, g_n$ and load factor converges to $z = 1$.

**Case 1:** Load factor $z < 1$. Here the link is underloaded, hence due to the first term $ACR(i)/z$, all the rates increase. If $n = 0$, i.e. all the sessions across this link are bottlenecked elsewhere. In this case since there are no non-bottlenecked sources, the GW fair allocation is trivially achieved. Assume that $n \geq 1$, now because of the first term $SourceRate(i)/z$ (Algorithm A, step 8), the rates of non-bottlenecked sources increase. This continues till load factor reaches a value greater than or equal to one. Hence we have shown that if load factor is less than one, the rates increase till the load factor becomes greater than one.

**Case 2:** Load factor $z > 1$. In this case if the link is not getting its $ExcessFairshare$ then, its rate increases, which might further increase $z$. This continues till all the sessions achieve at least their $ExcessFairshare$. At this point the allocation rates are decreased proportional to $1/z$ due to the first term. As in the previous case the $z$ decreases, till it reaches a value of 1 or less.

From the above two cases it can be seen that load factor oscillates around one and converges to the value of one. Assume that load factor is $z = 1 + \epsilon$, then the number round trip times for it to converge to one is given by $log_{1+\epsilon} \mid S_l \mid$. Henceforth, in our analysis we assume that the network is near the steady state that is load factor is near one. This implies

$$\sum_{i=1}^{k} r_{bi} + \sum_{i=1}^{n} r_i = A_l$$

$$\sum_{i=1}^{n} r_i = A_l - A_b$$

Let $A_m = \sum_{i=1}^{n} \mu_i$ be the total allocation for MCRs of the non-bottlenecked sources. Define $\alpha_i = r_i - \mu_i$, then we have

$$\sum_{i=1}^{n} \alpha_i = A_l - A_b - A_m = A$$

We have to show that:

$$\alpha_i = \frac{w_i A}{\sum_{j=1}^{n} w_j}$$

**Case A:** $B = 0$, i.e., there are no bottleneck sources. From the Algorithm A, step 8, we have

$$\alpha_i = Max(ExcessFairshare(i), \alpha_i/z)$$

We observe that the behavior of this equation behaves like a differential equation in multiple variables [21]. The behavior is like that of successive values of root acquired in the Newton-Ralphson method for finding roots of a equation. Hence the above equation converges, and the stable values of $\alpha_i$ is given by:

$$\alpha_i = ExcessFairshare(i) = \frac{w_i AL(i)A}{\sum_{j=1}^{n} w_j AL(i)}$$

Since we have assumed greedy sources and no bottlenecks in this case, the activity level is one for all sessions. Hence,

$$\alpha_i = \frac{w_i A}{\sum_{j=1}^{n} w_j}$$

which is indeed the desired value for $\alpha_i$.

**Case B:** $B \neq 0$, i.e., there are some bottleneck sources. Let $\beta_i$ be the allocated rate for corresponding to $r_{bi}$. Let $w_{bi}$ be the weight for session $s_{bi}$, $W_b = \sum_{i=1}^{K} w_{bi} AL(bi)$, $W = \sum_{i=1}^{n} w_i$. We know that the equation for the rate allocation behaves as a stabilizing differential equation. In the steady state all the above terms such as $W$, $W_b$ and rates stabilize. For bottlenecked sources the current link calculates a rate $\beta_i$ which is greater than $r_{bi}$, otherwise the bottlenecked session would be bottlenecked at the current link. For non-bottlenecked source the rate at steady state is given by:

$$\alpha_i = \frac{w_i(A_l - A_m)}{W_b + W}$$

Since the link has an overload of one at steady state we have

$$\sum_{i=1}^{n} \alpha_i = A_l - A_m - A_b$$

which implies that

$$\frac{\sum_{i=1}^{n} w_i(A_l - A_m)}{W_b + W} = A_l - A_m - A_b$$

$$W_b = \frac{WA_b}{A_l - A_m - A_b}$$

Using the above value for $W_b$ we get:

$$\alpha_i = \frac{w_i(A_l - A_m)}{\frac{WA_b}{A_l - A_m - A_b} + W}$$

$$\alpha_i = \frac{w_i(A_l - A_m - A_b)}{W}$$

which is the desired values for the $\alpha_i$. Hence, the sessions bottlenecked at the link $l_b$ do indeed achieve the GW fairness. □

**Theorem 1** *Starting at any arbitrary state of the network, if only greedy sources and source bottlenecked sources are present the Algorithm A converges to GW fair allocation.*

**Proof:** The convergence of the distributed algorithm similar to the centralized algorithm. Assume that the centralized algorithm converges in $M$ iterations. At each iteration there are set of links $\mathcal{L}_i$ which are bottlenecked at the current iteration. $\cup_{i=1}^{M} \mathcal{L}_i = \mathcal{L}$.

Using lemma 1, we know that each link $l \in \mathcal{L}_i$ does indeed converge to the general fair allocation $\mathcal{G}_l$. The distributed algorithm converges in the above order of links until the whole network is stable and allocation is $\mathcal{G}$. The number of round trips taken to converge is bounded by $MO(\log S)$, since each link takes $O(\log S_l)$ round trips for convergence. □

---

[3]All our papers and ATM Forum contributions are available through http://www.cis.ohio-state.edu/~jain/